\documentclass[reprint,twocolumn,showpacs,showkeys]{revtex4-1}

\usepackage{latexsym,graphicx,graphics}
\usepackage{appendix}
\usepackage{amsmath,amssymb,epsfig}
\usepackage[utf8]{inputenc}
\usepackage[mathscr]{euscript}
\usepackage{float}
\usepackage{tabularx,ragged2e,booktabs}

\newcommand{\abs}[1]{\left| #1 \right|}

\newcommand{\bb}[1]{ \mbox{\boldmath$ #1$}}

\newcommand{\rv}{\bb r}
\newcommand{\rvp}{\bb{r'}}

\newcommand{\Eq}[1]{Eq.~(\ref{#1})}
\newcommand{\Eqs}[2]{Eqs.~(\ref{#1})--(\ref{#2})}

\newcommand{\unit}[1]{\bb{\hat{#1}}}

\begin{document}

\title{Modal and excitation asymmetries in magnetodielectric particle chains}

\author{Y. Mazor}

\author{Ben Z. Steinberg}
\email{yardenm@gmail.com, steinber@eng.tau.ac.il}
\affiliation{
School of Electrical Engineering, Tel-Aviv University, Ramat Aviv, Tel-Aviv 69978 Israel
}%

\date{\today}

\begin{abstract}
We study the properties of dipolar wave propagation in linear chains of isotropic particles with independent electric and magnetic response, embedded in vacuum. It is shown that the chain can support simultaneously right-handed modes (RHM) and left-handed modes (LHM) of transverse-polarization. The LHM are supported by the structure even if the chain's particles possess positive polarizabilities and no Bi-isotropy; the needed structural Bi-isotropy is provided by the propagator instead of by the particle's local properties. In contrast to the transverse modes in chains that consist of purely electric particles that are inherently RHM, the LHM dispersion lacks the light-line branch since the dipolar features are not aligned with the electric and magnetic fields of a right-handed plane-wave solution in free space. Furthermore, it is shown that the spatial width of the LHM is significantly smaller than that of the RHM. Excitation theory is developed, and it is shown that the chain possesses modal and excitation asymmetries that can be used to eliminate reflections from chain's termination.
\end{abstract}

\maketitle

\section{Introduction}
Wave propagation and modal analysis for linear arrays of electrically polarizable particles were studied in many publications \cite{Markel1,Viitanen,TretyakovChain,EnghetaChain,HadadSteinbergGreen,CapolinoChain}. In its most basic form--a linear array of isotropic electric particles--the structure supports two independent modes: transverse modes, where the dipole moments are polarized perpendicular to the chain axis, and longitudinal modes for which the dipoles are polarized along the chain axis. In addition, it was shown that for the transverse modes, there is always a portion of the dispersion curve that runs adjacent to the light-line. Modes associated with this part of the dispersion, termed as the light-line modes, are typically very wide and resemble a plane wave interacting very weakly with the particle chain \cite{EnghetaChain}, and are hardly excitable \cite{HadadSteinbergGreen}. If the inter-particle distance $d$ is much smaller than the surrounding free-space wavelength $\lambda$, then the typical modal width away from the light-line (i.e. at wavenumbers $\beta(\omega)\gg k=\omega/c$) is also much smaller than $\lambda$ and the mode decays exponentially away from the chain \cite{TretyakovChain, EnghetaChain}. Hence these structures are often called \emph{sub-diffraction chains} (SDC). SDC's were suggested as potential candidates for ultra-narrow optical waveguides, junctions, couplers \cite{Vitaliy1}, as one-way guiding structures and optical isolators \cite{HadadSteinbergPRL,MazorSteinberg_LongChir}, and as leaky-wave antennas \cite{Vitaliy2,HadadSteinbergOpEx}.

Electric-magnetic and bi-isotropic particles, characterized by both electric and magnetic response were also studied extensively. In most cases, the context of these studies was the reflection, transmission and absorption properties of planar arrays of such particles under external plane-wave excitation, both reciprocal \cite{Tretyakov1,Tretyakov2,TretyakovReview,TretyakovPRX2015} and non-reciprocal \cite{TretyakovOneWaySheets}. 3D arrays of scalar magneto-electric particles were studied in \cite{Capolino_MagnetoElectric3D}, where the full inter-particle electric and magnetic coupling has been taken into account and the effect of array packaging on the electromagnetic modes has been studied. Such 3D arrays were also studied in \cite{AluHomog1,AluHomog2} in the context of homogenization techniques, and it has been shown that the homogenized material may possess negative index properties even if the microscale inclusions are made of conventional material (e.g. spherical particles made of material with positive scalar $\epsilon$ and $\mu$). Magnetic crystals (either 1D or 2D, respectively) with various models and interaction schemes taking into account short range and/or long range inter-particle interactions, were also considered \cite{Mag1,Mag2}.

Not much attention was given to the guiding properties of magneto-electric particle arrays, either in 1D or 2D. In the present work, we investigate the microscopic modal properties of such particle chains in vacuum. It is shown that the chain can support simultaneously (i.e. at the same frequency) right-handed modes (RHM) and left-handed modes (LHM) of transverse-polarization. This is to contrast with the 3D arrays in \cite{AluHomog1,AluHomog2} where the left-handedness and right-handedness are mutually exclusive. The LHM are supported by the structure even if the chain's particles possess positive polarizabilities and no Bi-isotropy; the needed structural Bi-isotropy is provided by the propagator instead of the particle's local properties. In contrast to the regular transverse RHM in purely electric particle chains,  the transverse LHM dispersion lacks the light-line branch discussed above since their dipolar features are not aligned with the electromagnetic fields of a right-handed plane-wave solution in free space. We study the modal confinement of the LHM and RHM around the chain, and show that the spatial width of the LHM is significantly smaller than that of the RHM. This is intuitively expected since the surrounding free-space is inherently right-handed and therefore it is less ``susceptive'' to the LHM. Hence, the chains studied here may be better suited for dense packaging of photonic systems. We use the Z-transform method to study the chains's excitation, and show that it possesses
asymmetries that can be used to eliminate unwanted reflections from chain's termination, paving the way to use the chain as a new kind of leaky wave antenna.

The structure of the paper is as follows. The formulation, based on the discrete dipole approximation, is presented in Sec.~\ref{Form}. The transverse modes are discussed in Sec.~\ref{Sec_TransModes}, where the corresponding dispersions, the existence of LHM and RHM, and the chain's eigenstates asymmetries are explored. A rigorous excitation theory, based on the Z transform, is outlined in Sec.~\ref{Sec_Excit} where the asymmetric excitation is presented and discussed, and exploited to eliminate back-reflection from a chain's termination. Concluding remarks are provided in Sec.~\ref{Conc}.

\section{Formulation}\label{Form}
\begin{figure}[h]
\begin{center}
\noindent
  \includegraphics[width=7cm]{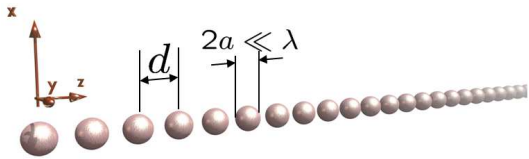}
  \caption{A linear chain of magnetic-electric particles. The particles are made of simple $\epsilon,\mu$ material, and no coupling between $\bb{E},\bb{H}$ occurs within the material.}\label{fig1}
\end{center}
\end{figure}

Consider the linear array of equally spaced isotropic particles shown in Fig.~\ref{fig1}.
 The inder-particle distance is $d$, and the particle diameter is much smaller than the wavelength $\lambda$ at which the structure is supposed to operate. Hence, the individual particle response to electromagnetic excitation is appropriately described by its polarizability matrix $\underline{\bb{\alpha}}$. We further assume that the particles possess electric and magnetic response only, with no bi-isotropic or bi-anisotropic properties (a simple example for such particles is a sphere made of conventional $\epsilon$ and $\mu$ material.)
 The corresponding $\underline{\bb{\alpha}}$ is a $6\times6$ matrix, describing the relation between the \emph{local} electromagnetic field $\bb{E},\bb{H}$ (the field at the particle's location in the absence of the particle) and the electric and magnetic dipole moments excited in the particle $\bb{p}$ and $\bb{m}$,
 \begin{subequations}
 \begin{equation}\label{eq1a}
 \bb{\Pi}\equiv\left(
 \begin{array}{c}
 \epsilon_0^{-1}\bb{p}\\
 \eta_0\bb{m}
 \end{array}\right)=\underline{\bb{\alpha}}\left(
\begin{array}{c}
 \bb{E}\\
 \eta_0\bb{H}
 \end{array}\right)
 \end{equation}
 where, in the quasi-static approximation
\begin{equation}\label{eq1b}
\underline{\bb{\alpha}}^{-1}=\left(
\begin{matrix}
\alpha_{ee}{\bf I}_3 & {\bf 0}_3 \\
{\bf 0}_3 & \alpha_{mm}{\bf I}_3
\end{matrix}
\right)^{-1}-\frac{2}{3}ik^3 {\bf I}_6,
\end{equation}
\end{subequations}
and where in $\underline{\bb{\alpha}}$ above we have used the unified notations defined in \cite{FemiusUnits}. For convenience they are summarized in Appendix \ref{App_Uni}. Unless otherwise stated, all bold-italic quantities here and henceforth represent column vectors, and underlined quantities represent matrices. ${\bf I}_\ell$ is a $\ell\times\ell$ identity matrix.
For a chain of identical particles, the chain dynamics is governed by the infinite matrix equation
\begin{equation}
\boldsymbol{\Pi}_n=\underline{\bb{\alpha}}\sum\limits_{\substack{m=-\infty \\ m\neq n}}^{\infty}
\underline{{\bf G}}_{\,n-m}\bb{\Pi}_m+\underline{\bb{\alpha}}\bb{F}_n^{\mbox{\tiny inc}}
\label{eq2}
\end{equation}
where $\bb{F}_n^{\mbox{\tiny inc}}=[\bb{E}^{\mbox{\tiny inc}}(\rv_n),\eta\bb{H}^{\mbox{\tiny inc}}(\rv_n)]^T$ is the incident field at the location of the $n$-th particle.
$\underline{{\bf G}}_{\,n}$ is a $6\times6$ matrix representation of the Green's dyadic, discussed in appendix \ref{App_GZ} and given by \Eqs{eqAppB_1}{eqAppB_2b} there. It consists of the four $3\times 3$ submatrices $\underline{{\bf G}}^{ee}_{\, n}$, $\underline{{\bf G}}^{em}_{\,n}$, $\underline{{\bf G}}^{me}_{\,n}=-\underline{{\bf G}}^{em}_{\,n}$, and $\underline{{\bf G}}^{mm}_{\,n}=\underline{{\bf G}}^{ee}_{\,n}$. The diagonal elements of $\underline{{\bf G}}^{em}_{\,n}$ vanish; this is a direct manifestation of the fact that the $\bb{H}$ ($\bb{E}$) field generated by the dipole $\bb{p}$ ($\bb{m}$) vanishes along the dipole axis. In fact, the structure of $\underline{{\bf G}}_{\,n}$ indicates that the chain supports four independent polarizations,

\begin{enumerate}

\item \label{LongE}
Longitudinal electric mode: $(\bb{E},\bb{p})=(\unit{z}E_z,\unit{z}p_z)$. The chain electrodynamics has been studied thoroughly. The modes were studied in e.g.~\cite{TretyakovChain,EnghetaChain}. Excitation (Green's function) theory has been developed in \cite{HadadSteinbergGreen}.

\item \label{LongH}
Longitudinal magnetic mode: $(\bb{H},\bb{m})=(\unit{z}H_z,\unit{z}m_z)$. Essentially the same as the point above.

\item \label{Mixed}
Transverse coupled (mixed) mode: $(E_x$,$p_x)$ and $(H_y$,$m_y)$. It will be formulated and studied below.

\item \label{Mixed2}
Transverse coupled (mixed) mode: $(E_y$,$p_y)$ and $(H_x$,$m_x)$. The same as in \ref{Mixed}.

\end{enumerate}

An important feature to note here is that there is a coupling between the magnetic and electric dipoles: an \emph{electric} dipole $\bb{p}=\underline{\bb{\alpha}}_{\,ee}\bb{E}^L$ excited by the local electric field $\bb{E}^L$ in a given particle, generates both $\bb{E}$ and $\bb{H}$ radiation fields. Then the $\bb{H}$ field excites \emph{magnetic} dipoles $\bb{m}=\underline{\bb{\alpha}}_{\,mm}\bb{H}^L$ in the neighboring particles. Hence, this coupling is \emph{non-local} in the sense that it is provided only by the field propagator and not by the properties of the chain's particles; the latter lack intrinsic bi-(an)isotropy.
In view of the above, we turn to study the transverse mixed modes.

\section{Transverse modes}\label{Sec_TransModes}

\begin{figure}[h]
\begin{center}
\noindent
\hspace*{-0.3cm}
  \includegraphics[width=8cm]{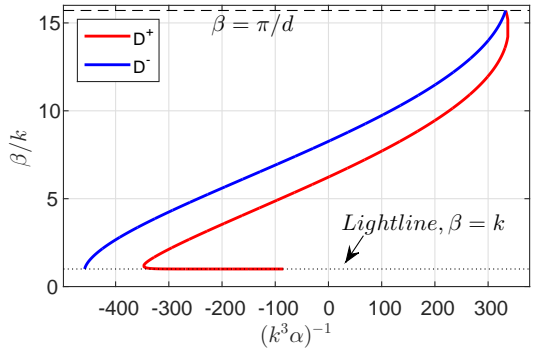}
  \caption{Dispersion for the transverse modes for $kd=0.2$, where $k$ is the free-space wavenumber. Left-handed modes--the $D^-$ dispersion curve--exist also for positive real polarizabilities.}\label{fig2}
\end{center}
\end{figure}

Since we are interested only with the transverse modes, we truncate Eq.~\ref{eq2} using it's 1,5 rows only, yielding
\begin{subequations}
\begin{equation}\label{eqAs1a}
\boldsymbol{\Pi}_{T,n}=\underline{\bb{\alpha}}_T\sum_{m\neq n}
\underline{{\bf G}}_{T,\,n-m}\bb{\Pi}_{T,m}+\underline{\bb{\alpha}_T}\bb{F}_{T,n}^{\mbox{\tiny inc}}
\end{equation}
where
\begin{equation}\label{eqAs1b}
\underline{{\bf G}}_{T,n}=\left(
\begin{matrix}
G^{ee}_{xx,\,n} & G^{em}_{xy,\,n}\\
G^{me}_{yx,\,n} & G^{mm}_{yy,\,n}
\end{matrix}\right)=\left(
\begin{matrix}
G^{ee}_{xx,\,n} & G^{em}_{xy,\,n}\\
G^{em}_{xy,\,n} & G^{ee}_{xx,\,n}
\end{matrix}\right)
\end{equation}
\end{subequations}
where we have used the identities $G^{ee}_{xx,\,n}=G^{mm}_{yy,\,n}$ and $G^{em}_{xy,\,n}=G^{me}_{yx,\,n}$ [see \Eqs{eqAppB_1}{eqAppB_3}]. We have also truncated $\boldsymbol{\alpha}$ in the same manner, $\underline{\bb{\alpha}}_{\,T}=\mbox{diag}(\alpha_{ee},\alpha_{mm})$. The modes supported by the structure are the solutions of \Eq{eq2} with no external forcing ($\bb{F}_n^{\mbox{\tiny inc}}=0\,\forall \, n$). Using Floquet's theorem with ${\bf{\Pi}}_T=[\epsilon_0^{-1}p_x,\eta_0m_y]^T$
\begin{equation}
\boldsymbol{\Pi}_{T,n}=\boldsymbol{\Pi}_{T,0}e^{i\beta nd}
\label{eq3}
\end{equation}
and the dynamic equation reduces to the $2\times 2$ system
\begin{subequations}
\begin{equation}
\underline{\bf{M}}(\alpha,Z){\bf \Pi}_{T,0}=0
\label{eq5a}
\end{equation}
with
 \begin{equation}\label{eq5bb}
 \underline{\bf{M}}(\alpha,Z)=\left[(k^3\underline{\bb{\alpha}}_{\, T})^{-1}-\left(
 \begin{matrix}
A_T(Z) & B(Z) \\
B(Z) & A_T(Z)
\end{matrix}\right)\right]
\end{equation}
\end{subequations}
$A_T(Z),B(Z)$ contain all the needed summations, given in \Eqs{eqAppB_5a}{eqAppB_5e} in Appendix \ref{App_GZ} with $Z=e^{-i\beta d}$. They incorporate all dipole-dipole interactions in this system, both short and long range. This physical fact is manifested mathematically by the dependence of $A_T(Z),B(Z)$ on the polylogarithm functions as detailed in the appendix, rather then on an inverse polynomial.
${\bf \Pi}_{T\,0}$ is the two-elements column vector $(\epsilon_0^{-1}p_{0x},\eta_0m_{0y})^{T}$. To obtain the dispersion for guided modes, we look the values of \emph{real} $\beta$  in the domain $\beta>k=\omega/c$ (outside of the light-cone) for which the determinant of \Eq{eq5a} vanishes.
For lossless particles, this condition guarantees the existence of solutions to the dispersion equation with $\Im[{\beta}]=0$. Using the properties pointed out by \Eqs{eqAppB_6a}{eqAppB_8b} in Appendix \ref{App_GZ}, we note that the imaginary part of $A_T$ cancels out with the $\bb{\alpha}_T^{-1}$ radiation damping factor (see also analysis in \cite{EnghetaChain}), and that in this domain $B$ is pure real.
The dispersion equation can then be simplified into
\begin{equation}
\begin{split}
& \left[(k^3\alpha_{ee})^{-1}-\Re\left\{A_T\right\} \right]\\
&\times \left[(k^3\alpha_{mm})^{-1}-\Re\left\{A_T\right\} \right]-B^2=0
\end{split}
\label{eq6}
\end{equation}
For simplicity we choose to focus on a case for which the particles are \emph{balanced}:  $\alpha_{ee}=\alpha_{mm}=\alpha$. Balanced particles have been considered for meta-surfaces applications in many previous publications - see Refs. \cite{Tretyakov1,Tretyakov2,TretyakovReview,TretyakovPRX2015,TretyakovOneWaySheets}. The realization of particles having both electric and magnetic response is possible even if we use simple dielectric materials. Dielectric spheres possess both electric and magnetic dipole resonances in positive values of $\epsilon$ which can be used as a simple platform to implement such systems, as reported in \cite{SihvolaSphere}. Tuning the different electric and magnetic dipole reosnances is possible via various geometrical transformations of the inclusions, as can be seen in \cite{Campione_Balanced}, and can be utilized in the design of balanced particles. This choice of particle simplifies our analysis and allows further reduction of the dispersion into two simple and distinct branches
\begin{equation}
(k^3\alpha)^{-1}=\Re\left\{A_T\right\}\pm B
\label{eq7}
\end{equation}
termed accordingly as $D^+$ and $D^-$. The dispersions are shown in Fig.~\ref{fig2}. The inter-particle distance is chosen such that $kd=0.2$. The dispersion is shown for positive $\beta$ values, and since the system is symmetric we will have the mirror image for negative $\beta$ (albeit the roles of $D^+$ and $D^-$ are switched).

Formally, there are four eigenvector solutions to \Eq{eq5a}. Examining \Eq{eq5a} together with the dispersion condition in \Eq{eq7}, one finds that there are two doubly-degenerate modes (DDM). Each DDM consists of two modes with the same dispersion, as expected from the symmetry of the chain in the $x,y$ plane. The dipole structure of each DDM is given by [note the definition of the vector ${\bf \Pi}$ in \Eq{eq1a}]
\begin{subequations}
\begin{equation}\label{eq8a}
\bb{p}_0=(p_{0x},0)^T,\,\bb{m}_0=(0,\gamma c p_{0x})^T
\end{equation}
or
\begin{equation}\label{eq8b}
\bb{p}_0=(0,p_{0y})^T,\,\bb{m}_0=(-\gamma c p_{0y},0)^T
\end{equation}
\end{subequations}
where $\gamma$ is given by
\begin{equation}
\gamma=\frac{B}{(k^3\alpha)^{-1}-\Re\left\{A_T\right\}}=\pm 1\quad\mbox{for } D^\pm.
\label{eq9}
\end{equation}
This degenerate mode is in fact the independent polarizations described in points \ref{Mixed} and \ref{Mixed2} in Sec.~\ref{Form}. Clearly, there are also two possible DDM's, the difference between them stems from the different choices of $\gamma$ above.

\subsection{Right-handed and left-handed modes}

From \Eqs{eq8a}{eq9} it follows that
\begin{equation}\label{eq10}
\bb{p}_0\times\bb{m}_0^*=\unit{z}\gamma c\abs{\bb{p}_0}^2.
\end{equation}
 Therefore along the dispersion branch $D^+$ ($\gamma=1$) the mode possesses a conventional right-hand (RH) structure. However, along $D^-$, where $\gamma=-1$, the mode is \emph{left-handed} (LH). As pointed out above, the corresponding dispersion curves are shown in Fig.~\ref{fig2}, and it is evident that LH modes can exist also for positive polarizability.

 It is interesting to note that the RH mode dispersion ($D^+$ in Fig.~\ref{fig2}) possesses the well known light-line branch that is external and adjacent to the light-line cone. This is very similar to the light-line transverse mode observed previously in e.g. Ref. \cite{EnghetaChain}. The corresponding electromagnetic wave interacts very weakly with the chain. It is spatially wide and possesses a plane-wave like structure. The mode dispersion is $\beta\approx k=\omega/c$, very close to that of a plane-wave.

To contrast, the LH mode possesses the $D^-$ dispersion curve. Due to its left-handedness it cannot be matched to the electromagnetic field supported by the surrounding vacuum. Hence
for this dispersion the light-line branch does not exist as can also be seen in Fig.~\ref{fig2}. We emphasize again that the left-handedness exists even when both polarizabilities (electric and magnetic) are positive. This regime does not exist in regular, electric particle chains and is the result of the field propagator coupling between $\bb{p}$ and $\bb{m}$.

\begin{figure}[h]
\begin{center}
\noindent
\hspace*{-0.3cm}
  \includegraphics[width=8cm]{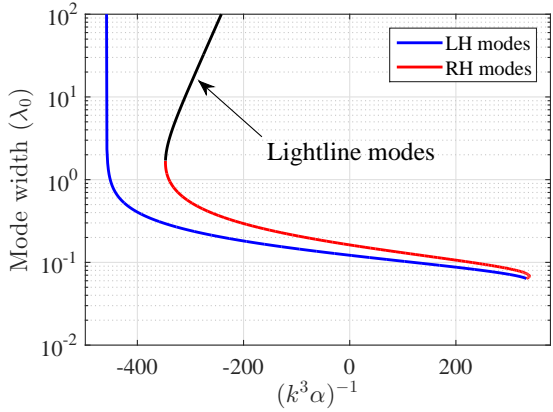}
  \caption{Spatial width of the LHMs and RHMs under guiding conditions, with the same parameters as in Fig.~\ref{fig2}.}\label{fig3}
\end{center}
\end{figure}

It is instructive to examine the spatial width of the electromagnetic field around the chain associated with each of these modes. The electric field is given by
\begin{equation}\label{eq11}
\bb{E}(\rv)=\sum_{n=-\infty}^{\infty}\left[
\underline{\bf G}^{ee}(\rv,\rv_n),\, \underline{\bf G}^{em}(\rv,\rv_n)
\right]
\bb{\Pi}_n
\end{equation}
where $\rv_n=\unit{z}nd$ is the location of the $n$-th particle, $\rv$ is any location off the chain particles, and where $\bb{p}_n,\bb{m}_n$ in $\bb{\Pi}_n$ satisfy the eigen-vector conditions of \Eqs{eq8a}{eq9}. This series can be evaluated with the aid of the Poisson summation formula. To estimate the mode width, it is sufficient to observe the leading term in the formula, that yields for $\bb{p}_0=\unit{x}p_{0x}$ and for $\rv=\unit{y}y+\unit{z}z$
\begin{equation}\label{eq12}
\begin{split}
E_x & = ip_{0x}\,\frac{k^2\, e^{i\beta z}}{4d}\\
    & \times\left[
\left(1+\gamma\frac{\beta}{k}\right)H_0^{(1)}(i\zeta y)-\frac{i\zeta}{k^2y}H_0^{(1)}(i\zeta y)\right]
\end{split}
\end{equation}
where $\zeta=\sqrt{\beta^2-k^2}$. The electric field of the LHM or RHM is obtained by using the corresponding $\beta$ and $\gamma$. Recall that for the guided mode $\beta>k$ so $\zeta$ is real. For large $y$ the expression above decays essentially as $e^{-\zeta y}$ away from the chain. Hence the characteristic width of the mode is $\zeta^{-1}$. From the dispersion curves of the LHM and RHM shown in Fig.~\ref{fig2} it is seen that always $\beta_{\mbox{\tiny LHM}}>\beta_{\mbox{\tiny RHM}}$. Hence, the LH mode is always more confined to the chain. This is intuitively expected, since the surrounding vacuum is a right-handed environment and hence is more susceptive to RHM than to LHM. Figure \ref{fig3} shows the characteristic widths of the LHM and RHM for $kd=0.2$. It is seen that the LHM confinement is better.

\subsection{Asymmetries of the chain's eigenstates}\label{sec_As}

\begin{figure}[h]
\begin{center}
\noindent
\hspace*{-0.1cm}
  \includegraphics[width=7.5cm]{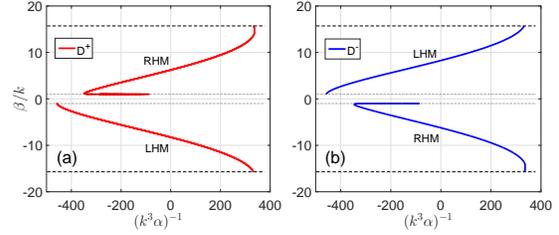}
  \caption{The dispersions of the diagonalized problem in \Eqs{eqAs3a}{eqAs3c}. (a) $q=1$ yields the $D^+$ curve only, and (b) $q=2$ yields the $D^-$ curve only. }\label{fig4}
\end{center}
\end{figure}

When our chain dynamics is expressed in terms of its eigenstates, a significant asymmetric behavior is exposed. This asymmetry stems from the nature of the LHM, RHM, and from the specific way they may couple to each other. Let us consider the propagator ${{\bf G}}_{T}$ in \Eq{eqAs1b}. This symmetric matrix has two distinct eigenvectors
\begin{equation}\label{eqAs2}
\bb{v}_1=\left(
\begin{matrix} 1\\ 1\end{matrix}\right),\quad
\bb{v}_2=\left(
\begin{matrix} 1\\ -1\end{matrix}\right)
\end{equation}
For simplicity, we again assume a balanced particle $\alpha_{ee}=\alpha_{mm}=\alpha$. The system in \Eq{eqAs1a} can now be \emph{diagonalized} by using the transformation matrix $\bb{T}=(\bb{v}_1,\bb{v}_2)$. The result is two \emph{decoupled} and scalar difference equations
\begin{subequations}
\begin{equation}\label{eqAs3a}
\Pi_{n}^{(q)}=\alpha\sum_{m\neq n}
G_{n-m}^{(q)}\Pi_{m}^{(q)}+\alpha F_n^{(q)}
\end{equation}
where $q=1,2$ and
\begin{equation}\label{eqAs3b}
G_n^{\left(\substack{1 \\ 2}\right)}=G^{ee}_{xx,\,n} \pm G^{em}_{xy,\,n}
\end{equation}
and where
\begin{equation}\label{eqAs3c}
\Pi_n^{(q)}=\bb{v}_{q}^T\boldsymbol{\Pi}_{T,n},\quad F_n^{(q)}=\bb{v}_{q}^T\bb{F}_{T,n}^{\mbox{\tiny inc}}.
\end{equation}
\end{subequations}
The two problems $q=1$ and $q=2$ are completely decoupled, each possesses its own dispersion. Furthermore, since $\bb{v}_1$ is the eigenmode associated with $q=1$, it determines the mode structure for \emph{both propagation directions}. Hence, this mode is RH for propagations towards $+z$, and LH for propagation towards $-z$. The case of $q=2$ is reversed. In fact, the dispersion for $q=1$ is given by the roots of \Eq{eq7} with the $+$ sign, i.e.~it is nothing but $D^+$ \emph{only}. Likewise, the dispersion for $q=2$ is given by the roots of \Eq{eq7} with the $-$ sign, hence it is nothing but $D^-$ \emph{only}. These dispersions are shown in Fig.~\ref{fig4}

Clearly, each mode's dispersion is un-even with respect to the propagation wavenumber $\beta$. At frequencies corresponding to $-450<(k^3\alpha)^{-1}<-350$ it seems that there is only one-way propagation for each of the modes. Nevertheless, the chain itself is reciprocal. This apparent contradiction with the Lorentz reciprocity theorem stems from the fact that each of the de-coupled formulations in \Eqs{eqAs3a}{eqAs3c} represent only a part, or a projection, of the chain dynamics, and does not expose the subtleties associated with the problem \emph{excitation}. This is further discussed in Appendix \ref{App_Recip}.

This type of decomposition is useful in determining the response to a general point source $(p,m)^T$. We decompose the source using \Eq{eqAs3c}, then calculate the response to each of the $1,2$ components independantly. The total response is given by the superposition of these. Naturally, if both $p$ and $m$ are present, we will still obtain an asymmetric response.

\begin{figure}[h]
\begin{center}
\noindent
\hspace*{-0.1cm}
  \includegraphics[width=7.5cm]{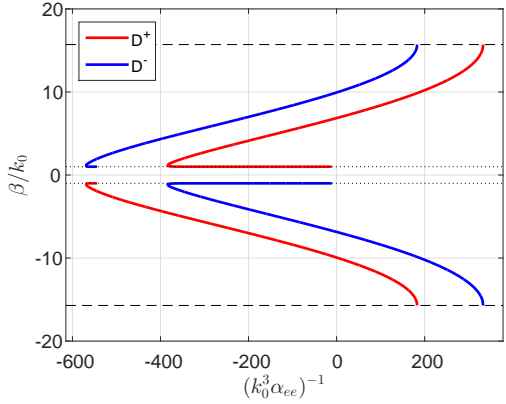}
  \caption{Dispersion for a chain of unbalanced particles for $\delta=100$.}\label{fig_unbalanced}
\end{center}
\end{figure}

 Our analysis can also be applied to the case of unbalanced particles. This would lead to eigenmodes which are not purely RHM or LHM, but rather some mixture of the two. Further, this will cause the dispersion of both eigenstates to have lightlines (a situation that in some sense resembles excitation by a general source, as discussed). To elucidate this, we choose a simple model where $\alpha_{mm}^{-1}=\alpha_{ee}^{-1}+\delta$, which creates not only unequal $\alpha_{ee}$ and $\alpha_{mm}$, but also a deviation in their resonance frequencies ($\alpha_{ee}^{-1}=0$ or $\alpha_{mm}^{-1}=0$). For this case, \Eq{eq6} can then be written as
\begin{equation}
\alpha^{-1}_{ee}=A_T+\frac{-\delta\pm\sqrt{ \delta^2+4B^2 }}{2}
\label{unbalanced1}
\end{equation}
Both branches are shown in Fig.~\ref{fig_unbalanced}. As we see, there is a lightline branch for both $D^+$ and $D^-$, yet the strong asymmetry is maintained, and all further conclusions still apply, with the proper adjustments to the eigenmodes.
\section{Excitation theory}\label{Sec_Excit}

The excitation properties of each of the modes discussed in previous sections, as well as of other wave-constituents that may be supported by the chain, are fully exposed by the chain's Green's function matrix $\underline{\mathbb{G}}_{\,n}$, defined as the response to a $\delta$-dyad. Continuing our  interest in the mixed modes, we use the truncated dynamics equation \Eq{eqAs1a}. Under this truncation $\underline{\mathbb{G}}_{\,n}$ is a $2\times 2$ matrix sequence whose first [second] column describes the dipole chain response $(p_x,m_y)^T$ to $\bb{F}_n^{\mbox{\tiny inc}}=\delta_n (E_x^{\mbox{\tiny inc}},0)^T$ [$\delta_n(0, \eta H_y^{\mbox{\tiny inc}})^T$]. The response to any incident field $\bb{F}_n^{\mbox{\tiny inc}}=(E_{nx}^{\mbox{\tiny inc}},\eta H_{ny}^{\mbox{\tiny inc}})^T$ is then obviously obtained via the discrete convolution of the latter with $\underline{\mathbb{G}}_{\,n}$.

To derive $\underline{\mathbb{G}}_{\,n}$ we use the double-sided $Z$ transform commonly applied to analyze discrete systems and difference equations \cite{ZTBook1,ZTBook2}, and has been applied also to discrete electromagnetic systems in \cite{HadadSteinbergGreen,ZT_WasylWasyl,ZT_SIAM,CapolinoZTWH}. The present study follows essentially the same steps as in \cite{HadadSteinbergGreen}. For convenience, some details are provided in Appendix \ref{App_ZT}.

The chain dynamics in \Eq{eqAs1a} can be presented as the discrete convolution formula
\begin{subequations}
\begin{equation}\label{eq13a}
\sum_{m=-\infty}^\infty \underline{\bf D}_{\, n-m}\bb{\Pi}_m=\bb{F}_n^{\mbox{\tiny inc}}
\end{equation}
where
\begin{equation}\label{eq13b}
\underline{\bf D}_{\, n}=\left\{
\begin{array}{ll}
\underline{\bb{\alpha}}_T^{-1}\quad & n=0\\
-\underline{\bf G}_{T, n} \quad    & n\ne 0.
\end{array}\right.
\end{equation}
\end{subequations}
We apply the $Z$ transform on the equation above, using the notations
\begin{equation}\label{eq14}
\underline{\bf\hat{D}}(Z)=\sum_{n=-\infty}^\infty\underline{\bf D}_{\, n}Z^{-n}
\end{equation}
and similarly to all other involved quantities. With the convolution theorem \Eq{eq13a} reduces to
\begin{equation}\label{eq15}
\bb{\hat{\Pi}}(Z)=\left[ \underline{\bf\hat{D}}(Z)\right]^{-1}\bb{\hat{F}}(Z).
\end{equation}
Hence $\underline{\mathbb{G}}_{\, n}$ is given by the inverse Z transform (IZT)
\begin{subequations}
\begin{equation}\label{eq16a}
\underline{\mathbb{G}}_{\, n}=
\frac{1}{2\pi i}\oint_{C_\pm} \underline{\hat{\mathbb{G}}}(Z)\, Z^{n-1} dZ,
\end{equation}
where,
\begin{equation}\label{eq16b}
\underline{\hat{\mathbb{G}}}(Z)=\left[ \underline{\bf\hat{D}}(Z)\right]^{-1}
\end{equation}
\end{subequations}
By applying the ZT to \Eq{eq13b} we obtain (see Appendix \ref{App_GZ})
\begin{equation}\label{eq17}
\underline{\bf\hat{D}}(Z)=\underline{\bf{M}}(\alpha,Z)
\end{equation}
from \Eq{eq5bb}. The inverse of the above yields
\begin{equation}\label{eq18}
\begin{split}
& \underline{\hat{\mathbb{G}}}(Z)=
\frac{1}{|\underline{\bf\hat{D}}|}\times\\
& \left[
\begin{matrix}
(k^3\alpha)^{-1}-A_T(Z) & B(Z)\\
B(Z) & (k^3\alpha)^{-1}-A_T(Z) \end{matrix}\right]
\end{split}
\end{equation}
where we have again assumed the use of a balanced particle $\alpha_{ee}=\alpha_{mm}=\alpha$. While one may always apply the IZT on the last expression numerically and obtain the exact Green's function, a good physical insight is gained by examining the analytic properties of $\underline{\hat{\mathbb{G}}}(Z)$ in the complex $Z$ plane. Each and every singularity represents a distinct wave phenomena, and it's excitation is nothing but the corresponding residue. We note that $A_T(Z),B(Z)$ mainly consist of summation of polylogarithm functions $Li_n(e^{ikd}Z^{\pm1})$, see \Eqs{eqAppB_5b}{eqAppB_6a} (Appendix \ref{App_GZ}). Since $Li_1(z)=-\ln (1-z)$ and $Li_s'(z)=z^{-1}Li_{s-1}(z)$, then $\forall\,\, n>0$ $Li_n$ inherit the branch point and branch cuts singularities of $\ln (1-z)$ at $z=1$, creating Riemann sheets of infinite multiplicity. In the principal Riemann sheet $R0$, $A_T(Z),B(Z)$ and $\underline{\hat{\mathbb{G}}}(Z)$ possess two branch points at $Z_{b\, 1,2}=e^{\pm ikd}$ with one cut that emerges from $e^{-ikd}$ and extends to infinity, and a second cut that emerges from $e^{ikd}$ and extends to the origin. Further details regarding the discontinuity across the cuts can be found in \cite{PolylogBook} or in the Appendix of Ref. \cite{HadadSteinbergGreen}.

Since the polylogarithm function has no poles and no zeros, the pole singularities of $\underline{\hat{\mathbb{G}}}(Z)$ are only due to the zeros of its denominator, namely $|\underline{\bf\hat{D}}(Z_p)|=0$. Due to the specific structure of $A_T(Z),B(Z)$ [note \Eq{eqAppB_5e}] it follows that all the singular points must satisfy \emph{inversion symmetry}. That is, if $Z_{p\,\ell}$ is a pole, then $Z_{p\,\ell'}=1/Z_{p\,\ell}$ must also be a pole. This is quite general, and holds also for the branch points and cuts discussed above--a consequence of the chain's reciprocity. Furthermore, the pole equation is in fact a slight generalization of \Eq{eq7} with $A_T=A_T(Z)$ and $B=B(Z)$,
\begin{subequations}
\begin{equation}\label{eq19a}
(k^3\alpha)^{-1}=A_T(Z_p)\pm B(Z_p)\,\Leftrightarrow \, D^\pm.
\end{equation}
associating the pole singularities with the chain modes. There are six roots to this equation, providing six poles. The modes' dispersion are the solution of
\begin{equation}\label{eq19b}
e^{i\beta d}=Z_{p\ell}(\alpha),\quad \ell=1,\ldots 6.
\end{equation}
\end{subequations}
The curves presented in Fig.~\ref{fig2} correspond to poles that reside exactly on the unit circle in the complex $Z$ plane and consequently admit real $\beta$ in the equation above, and then the radiation damping in $\alpha$ precisely cancels out with $\Im\{A_T\}$. In the general case, however, depending on the values of $\alpha$ the poles $Z_p$ may reside off the unit circle, yielding \emph{leaky waves} (and also lossy waves, if $\alpha$ consists of material loss). Following the discussion of the IZT in Appendix \ref{App_ZT} poles for which $\abs{Z_p}\gtrless 1$ represent wave constituents of $\underline{\mathbb{G}}_{\, n}$ that contribute to $n\lessgtr 0$, respectively - see Fig.~\ref{fig1App_ZT} and explanation therein.
Finally, we note that since the guided modes' poles in the lossless chain satisfy $\abs{Z_p}=1$, their classification as singularities that contribute to $n>0$ ($n<0$) encircled by $C_+$ in Fig.~\ref{fig1App_ZT} (encircled by $C_-$) cannot be done according to their locations inside (outside) the unit circle. In this case the classification is done according to the group velocity $v_g=\partial\omega/\partial\beta=ide^{i\beta d}\left[Z_p'(\alpha)\,\partial\alpha/\partial\omega\right]^{-1}$. Alternatively one may observe the shift of $Z_p$ off the unit circle when loss is added to the system, and classify accordingly.

\begin{figure}[h]
\begin{center}
\noindent
\hspace*{-0.3cm}
  \includegraphics[width=8cm]{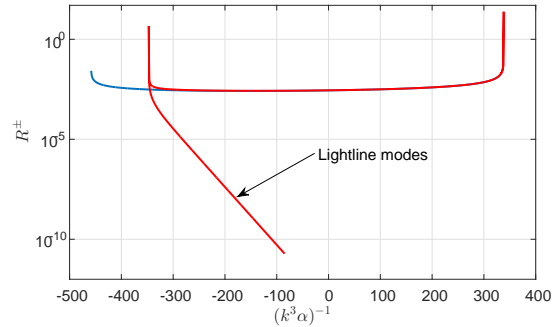}
  \caption{Residue (excitation magnitude) of the LHM and RHM under guiding conditions, with the same parameters as in Fig.~\ref{fig2}.}\label{fig5}
\end{center}
\end{figure}

Since there are six poles and they satisfy inversion symmetry, we denote by $Z_{p\,1,2,3}$ those that contribute to $n>0$ (with $|Z_{p\,1,2,3}|\le 1$) and by $Z_{p\, 4,5,6}=1/Z_{p\, 1,2,3}$ those that contribute to $n<0$. Likewise, the branch cut that emerges from $Z_{b\, 1}=e^{ikd}$ ($Z_{b\, 2}=e^{-ikd}$) and extends to the origin (to infinity) contributes to $n>0$ ($n<0$). Thus, by applying the residue theorem to \Eq{eq16a} for $n>0$ we find
\begin{subequations}
\begin{equation}\label{eq20a}
\underline{\mathbb{G}}_{\, n}=\sum_{\ell=1}^3\underline{\mathbb{G}}^{p\ell}_{\, n}+
\underline{\mathbb{G}}^{b2}_{\, n}
\end{equation}
where $\underline{\mathbb{G}}^{b2}_{\, n}$ is the result of an integration around the branch point no. 2 and the corresponding cut, and $\underline{\mathbb{G}}^{p\ell}_{\, n}$ is the residue of the $\ell$-th pole. It is given by
\begin{equation}\label{eq20b}
\underline{\mathbb{G}}^{p\ell}_{\, n}=
\frac{
-\left(
\begin{matrix}
\pm 1 & 1\\
1 & \pm 1
\end{matrix}\right)}
{2\left[\pm A_T'(Z_{p\ell})+ B'(Z_{p\ell})\right]}\,Z_{p\ell}^{n-1}
\end{equation}
\end{subequations}
where a prime indicates a derivative with respect to the argument, and where we have used the fact that the poles satisfy \Eq{eq19a}. Using the corresponding expressions in appendix \ref{App_GZ} and \Eq{eqAppB_9} there we finally find for the pole residues
\begin{subequations}
\begin{equation}\label{eq21a}
\underline{\mathbb{G}}^{p\ell}_{\, n}=
R^\pm
\frac{kd}{2}
\left(
\begin{matrix}
\pm 1 & 1\\
1 & \pm 1
\end{matrix}\right)\, Z_{p\ell}^n,\quad\mbox{for }D^{\pm}
\end{equation}
where $R^\pm$ is the amplitude
\begin{equation}\label{eq21b}
\begin{split}
R^\pm &=\bigl[ 2Li_0(e^{ikd}Z_{p\ell}^{\mp1})\\
      &+\frac{2i}{kd}Li_1(e^{ikd}Z_{p\ell}^{\mp1})\mp f_2^-(kd,Z_{p\ell})\bigr]^{-1}
      \end{split}
\end{equation}
\end{subequations}
Recall that for $n>0$ the $D^+$ dispersion represents RH modes, while the $D^-$ represents LH modes. For $n<0$ their roles interchange. As the pole approaches the branch-point $Z_{p\ell}\rightarrow Z_{b\,2}=e^{ikd}$ we have $R^+\rightarrow 0$. This is a manifestation of the fact that the conventional light-line mode is hardly excited as has already been observed in previous studies \cite{EnghetaChain,HadadSteinbergGreen}. However, it is interesting to note that $R^-$ stays finite in this limit. Hence, in contrast to the RH modes, the LH mode is practically excitable when it approaches the light-line. Figure \ref{fig5} shows the excitation amplitude (residues) of the RH and LH modes.

\subsection{Mode selectivity and LHM Excitation}

An obvious way to excite the LH mode in the chain is to place a dipole in the chain axis, whose frequency is in the range where the RH mode doesnot possess real $\beta$. Referring to Figs \ref{fig2} and \ref{fig4}, this corresponds to $-450\le (k^3\alpha)^{-1}\le -350$. Clearly, the particle's dipole response is negative there. A more interesting case is to work in the domain of positive $\alpha$, but as seen in Fig.~\ref{fig2} in this domain both LH and RH modes can exist. An insight into the excitation possibilities is provided by \Eqs{eq21a}{eq21b}. We emphasize that the upper (+) sign and the lower (-) sign correspond, respectively, to the RHMs and LHMs in the $n>0$ domain. For $n<0$ their roles interchange. In light of this fact and in light of the analysis presented in Sec.~\ref{sec_As}, we now examine the modes excitation when a Huygens source is placed in the chain. In our context a Huygens source is a source composed of mutually orthogonal electric and magnetic dipoles, that share the same center, oscillate at the same frequency, and possess the same phase. The two options are summarized in table \ref{table1}.
\begin{table}[H]
  \begin{tabular}{ p{2.5cm} || p{2cm} | p{2cm} }
    \hline
    $(\epsilon_0^{-1}p_x,\eta_0m_y)=$ & (1,1) & (1,-1) \\ \hline\hline
    Free-space radiation & $\unit{z}$ & $-\unit{z}$ \\ \hline
    Chain excitation & RH to $n>0$ LH to $n<0$ & LH to $n>0$ RH to $n<0$ \\
    \hline
    Chain eigenstates & $q=1$ only & $q=2$ only\\
    \hline
    Mode dispersion & Fig.~\ref{fig4}a & Fig.~\ref{fig4}b \\
    \hline
  \end{tabular}
  \caption{Summary of Huygens source excitation properties}
  \label{table1}
\end{table}
The observation summarized in Table \ref{table1} is a direct manifestation of the chain's \emph{left-handedness}: it provides an \emph{inversion of the radiation properties of Huygens sources}, including the wave handedness. If losses are present in the particles, of course the guided wave would be attenuated due to dissipation. However, The essential [a]symmetries are maintained, and can be exploited, in the same manner. This fact is discussed and demonstrated for example in \cite{MazorSteinberg_LongChir}.

\begin{figure}[h]
\begin{center}
\noindent
\hspace*{-0.3cm}
  \includegraphics[width=8cm]{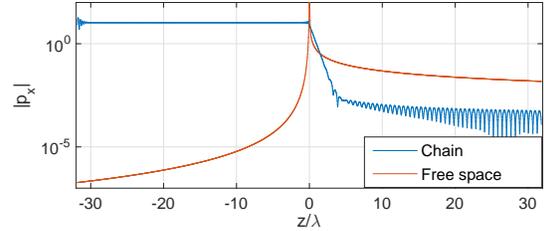}
  \caption{Response to a $(1,1)$ Huygens source. The chain inverts its properties.}\label{fig6}
\end{center}
\end{figure}

Finally, we note that by placing a source of type $(1,1)$ above, in a chain with $(k^3\alpha)^{-1}=-400$ the RHM is leaky and does not propagate in the chain. Then the guided mode excited by the source is only the LHM, and it propagates only into the $n<0$ domain. This case is shown in Fig.~\ref{fig6}.

\subsection{Excitation of finite chains}

\begin{figure}[h]
\begin{center}
\noindent
\hspace*{-0.3cm}
  \includegraphics[width=8cm]{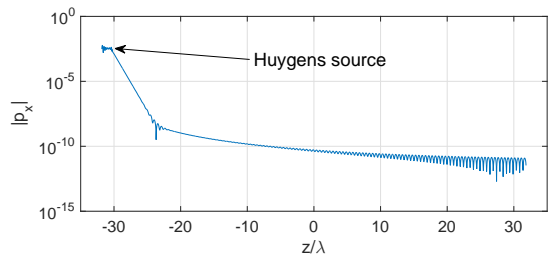}
  \caption{Response to a Huygens source $(1,1)$ with $(k^3\alpha)^{-1}=-400$ located near the termination of a finite chain. There is no reflection from the chain end.}\label{fig7}
\end{center}
\end{figure}

Recall the discussion in Sec.~\ref{sec_As}. The diagonalization procedure and the ensuing decoupled formulations in \Eqs{eqAs3a}{eqAs3c} hold also for finite or semi-infinite chains. The only difference is in the summation limits, but not in the diagonalizing transformations. Hence, chain termination (e.g.~semi-infinite chain, extending from $n=0$ to infinity) does not cause any mode-mixing.
For a Huygens source of the type $q=1$ [i.e.~$(\epsilon_0^{-1}p_x,\eta_0m_y)=(1,1)$] located at some $n'\gg 1$, a LHM would propagate towards the chain termination at $n=0$. When this mode hits the termination, it can only be reflected into a RHM. Thus, if the frequency is in the domain $-450\le (k^3\alpha)^{-1}\le -350$, where the RHM is in fact a leaky wave, there would be no guided reflection. Figure \ref{fig7} shows the response of a finite chain for a Huygens source $(1,1)$ with $(k^3\alpha)^{-1}=-400$. The mode leaks to free space with practically zero reflection into the chain.

\section{Conclusion}\label{Conc}

In this work we have formulated the equations governing the dipolar wave propagation in magneto-dielectric particle chains. The modal properties were extracted. It was shown that these structures can support simultaneously left-handed and right-handed modes (LHM and RHM, respectively), and that the LHM are narrower then their right-handed counterparts. LHM dispersion present no lightline - a consequence of their poor matching to free-space field propagation. Modal asymmetry was studied, and the asymmetric excitation of the LHMs and RHMs by Huygens sources was demonstrated. Complete excitation theory was developed for the propagating and leaky modes, and an explicit form of the chain Green's function was obtained. It has been shown that the modal asymmetry can be exploited to eliminate guided parasitic reflections from the termination of finite or semi-infinite chains, thus enabling the use of these chains as matched leaky-wave antennas. For simplicity of derivations we have assumed throughout the work that the particle is balanced, i.e.~$\alpha_{ee}=\alpha_{mm}$. Balanced scalar and bi-anisotropic particles were used previously in a number of publications \cite{Tretyakov1,Tretyakov2,TretyakovReview,TretyakovPRX2015,TretyakovOneWaySheets}, and realization using dielectric material has been suggested in \cite{Campione_Balanced}. The same analysis techniques that are used here, can be applied also to explore the case of non-balanced (and even bi-anisotropic) particles.
\appendix

\section{Unified notations}\label{App_Uni}

 In SI units, the dipole moments $\bb{p}$ and $\bb{m}$ possess the physical dimensions of C$\times$m and A$\times\mbox{m}^2$, respectively (here C,m, and A denote Coulomb, meter, and Ampere respectively). Hence all the entries of the six-element column vector $\bb{\Pi}$ have units of $\mbox{Volt}\times\mbox{m}^2$. For an isotropic particle, we denote the scalar $\alpha_{ee}^{\mbox{\tiny SI}}$ ($\alpha_{mm}^{\mbox{\tiny SI}}$) as the polarizability in SI units representing the electric (magnetic) dipole response due to a unit electric (magnetic) local field. Then the entries in \Eq{eq1b} are given by
\begin{equation}\label{eqAppA_1}
\alpha_{ee}=\alpha_{ee}^{\mbox{\tiny SI}}/(4\pi\epsilon_0),\quad \alpha_{mm}=\alpha_{mm}^{\mbox{\tiny SI}}/(4\pi)
\end{equation}
and both possess the dimensions of $\mbox{m}^3$.
Finally, we note that $\alpha_{ee}$ and $\alpha_{mm}$ above are the static polarizabilities. The expression in \Eq{eq1b} provides the quasi-static approximation where the imaginary additive part accounts for the particles radiation damping.

\section{The Green's function dyad}\label{App_GZ}

Clearly, a time-varying electric dipole generates both $\bb{E}$ and $\bb{H}$ fields. The same holds for a time-varying magnetic dipole. If the particles possess only electric response then $\bb{H}$ can be ignored, as is usually the case in most of the previous studies of particle chains. However, in our case the particles are assumed to possess both electric and magnetic dipole responses, hence it is crucial to take into account both fields for each dipole. Towards this end, we use the expressions given e.g. in \cite{JacksonBook} and
define the $6\times 6$ dyadic Green's function matrix via the relation
\begin{equation}\label{eqAppB_1}
\begin{split}
4\pi &
\left(
\begin{matrix}
\bb{E}(\rv)\\
\eta\bb{H}(\rv)
\end{matrix}
\right)=\underline{\bf G}(\rv,\rvp)\,\bb{\Pi}=\\
&\\
=&
\left(
\begin{matrix}
\underline{\bf G}^{ee}(\rv,\rvp) & \underline{\bf G}^{em}(\rv,\rvp)\\
\underline{\bf G}^{me}(\rv,\rvp) & \underline{\bf G}^{mm}(\rv,\rvp)
\end{matrix}
\right)\,\bb{\Pi}
\end{split}
\end{equation}
where $\bb{E}(\rv),\bb{H}(\rv)$ are the fields at $\rv$ due to the electric and magnetic dipoles $\bb{p}$ and $\bb{m}$ located at $\rvp$, and where $\bb{\Pi}$ is defined in \Eq{eq1a}. Generally we have $\underline{\bf G}^{ee}=\underline{\bf G}^{mm}$, $\underline{\bf G}^{em}= -\underline{\bf G}^{me}$. Since the chain coincides with the $z$ axes, we are interested only in the case $(\rv,\rvp)=(nd\unit{z},n'd\unit{z})$. Then, these $3\times 3$ dyads can be rewritten as $\underline{\bf G}^{ee}(\rv,\rvp)\rightarrow \underline{\bf G}^{ee}_{\, n-n'}$ etc..., with
\begin{subequations}
\begin{equation}\label{eqAppB_2a}
\underline{\bf G}^{ee}_{\, n}=
\left[k^2\underline{\bf A}_{\,1}-\left(\frac{1}{(nd)^2}+\frac{ik}{\abs{n}d}\right)\underline{\bf A}_{\,2}\right]
\frac{e^{ikd\abs{n}}}{\abs{n}d}
\end{equation}

where $\underline{\bf A}_{\,1,2}$ are the matrices $\underline{\bf A}_{\, 1}=\mbox{diag}(1,1,0)$,
$\underline{\bf A}_{\, 2}=\mbox{diag}(-1,-1,2)$, and where
\begin{equation}\label{eqAppB_2b}
\begin{split}
\underline{\bf G}^{em}_{\,n}  = & -k^2\left(1+\frac{i}{\abs{n}kd}\right)\\
&\\
 & \times \left(
\begin{matrix}
0 & -1 & 0\\
1 & 0 & 0\\
0 & 0 & 0
\end{matrix}
\right)\,\frac{e^{ikd\abs{n}}}{\abs{n}d}\,\mbox{sgn}(n).
\end{split}
\end{equation}
\end{subequations}
The Green's dyad in \Eq{eq2} consists of the four $3\times 3$ matrices $\underline{\bf G}^{ee}_{\,n}$,
$\underline{\bf G}^{em}_{\,n}$, $\underline{\bf G}^{me}_{\,n}=-\underline{\bf G}^{em}_{\,n}$, and
$\underline{\bf G}^{mm}_{\,n}=\underline{\bf G}^{ee}_{\,n}$.

Anticipating the use of Floquet's theorem for modal analysis, and further the use of $Z$-transform to study the excitation properties of the chain, we now look for the matrix summation
\begin{equation}\label{eqAppB_3}
\underline{\bf\tilde{G}}(Z)=\sum\limits_{\substack{n=-\infty \\ n\neq 0}}^{\infty}\underline{\bf G}_{\,n}Z^{-n},
\end{equation}
where, for the application of Floquet's theorem one merely substitutes the special case $Z=e^{-i\beta d}$.
The $Z$-transform above can be expressed in terms of the polylogarithm functions $Li_s(z)$ \cite{PolylogBook} defined as
\begin{equation}\label{eqAppB_4}
Li_s(z)=\sum_{n=1}^\infty\frac{z^n}{n^s}\,\Rightarrow\, Li_s'(z)=z^{-1}Li_{s-1}(z).
\end{equation}
Strictly speaking, the sum converges only for $\Re z \le 1$, but it can be analytically continued into the entire complex $z$ plane by noting that $Li_0(z)=z/(1-z)$, $Li_1(z)=-\ln (1-z)$, and by integrating over the second identity in \Eq{eqAppB_4}. Further details and properties of the polylogarithm functions can be found in Ref. \cite{PolylogBook}.

With the use of $Li_s(\cdot)$ the $6\times 6$ matrix $\underline{\bf\tilde{G}}(Z)$ can be expressed as
\begin{subequations}

\begin{widetext}

\begin{equation}\label{eqAppB_5a}
\underline{\bf\tilde{G}}(Z)=k^3\left(
\begin{matrix}
A_T(Z) & 0 & 0 & 0 & B(Z) & 0 \\
0 & A_T(Z) & 0 & -B(Z) & 0 & 0 \\
0 & 0 & A_L(Z) & 0 & 0 & 0 \\
0 & -B(Z) & 0 & A_T(Z) & 0 & 0 \\
B(Z)  & 0 & 0 & 0 & A_T(Z) & 0 \\
0 & 0 & 0 & 0 & 0 & A_L(Z)
\end{matrix}
\right)
\end{equation}
\end{widetext}
where
\begin{eqnarray}
A_T(Z) &=& f_1^+(kd,Z)\nonumber\\
\label{eqAppB_5b}\\
       &+& if_2^+(kd,Z)-f_3^+(kd,Z)\hspace*{0.35in}\nonumber\\
       \nonumber\\
A_L(Z) &=& 2\left[ f_3^+(kd,Z)-if_2^+(kd,Z)\right]\label{eqAppB_5c}\\
\nonumber\\
B(Z)   &=& f_1^-(kd,Z)+if_2^-(kd,Z)\label{eqAppB_5d}
\end{eqnarray}
and where
\begin{equation}
f_s^{\pm}(x,Z) \equiv x^{-s}\left[Li_s\left(e^{ix}Z^{-1}\right)\pm Li_s\left(e^{ix}Z\right)\right]\label{eqAppB_5e}
\end{equation}
\end{subequations}

Some properties of the functions above are worth pointing out. First, we note that on the unit circle in the complex $Z$ plane, i.e.~for $Z=e^{i\theta}$ with $\theta$ real,
\begin{subequations}
\begin{equation}\label{eqAppB_6a}
Li_n(e^{\pm i\theta})=C_n(\theta)\pm iS_n(\theta)
\end{equation}
where $C_n(\theta)$ and $S_n(\theta)$ are the generalized Clausen functions,
\begin{eqnarray}
C_n(\theta) &=& \sum_{k=1}^\infty\frac{\cos(k\theta)}{k^n}\label{AppB_6b}\\
S_n(\theta) &=& \sum_{k=1}^\infty\frac{\sin(k\theta)}{k^n}\label{AppB_6c}.
\end{eqnarray}
\end{subequations}
These functions are real and $2\pi$-periodic. Furthermore, for $0\le \theta\le 2\pi$
\begin{subequations}
\begin{eqnarray}
S_1(\theta) &=& \frac{1}{2}(\pi-\theta) \label{eqAppB_7a}\\
C_2(\theta) &=& \frac{\pi^2}{6} +\frac{\theta}{4} (\theta-2\pi) \label{eqAppB_7b}\\
S_3(\theta) &=& \frac{\theta}{12}\left(2\pi^2-3\pi\theta+\theta^2\right). \label{eqAppB_7c}
\end{eqnarray}
\end{subequations}
Using these identities, it is straight-forward to show that for $Z=e^{i\theta}$ with $\theta$ real and $\theta>kd$,
\begin{subequations}
\begin{eqnarray}
\Im\left\{B(e^{i\theta})\right\} &=& 0\label{eqAppB_8a}\\
\Im\left\{A_T(e^{i\theta})\right\} &=& 2/3.\label{eqAppB_8b}
\end{eqnarray}
\end{subequations}
We note that the second equality has been pointed out already in \cite{EnghetaChain}.

Finally, we note that the derivatives of $f_s^\pm$ with respect to $Z$ are needed for residue estimation in the chain's Green function derivation. We have
\begin{equation}\label{eqAppB_9}
\frac{\partial}{\partial Z}\,f^\pm_s(x,Z)=-(xZ)^{-1}f^\mp_{s-1}(x,Z).
\end{equation}

\section{The double sided Z transform}\label{App_ZT}

\begin{figure}[ht]
  \includegraphics[width=7cm]{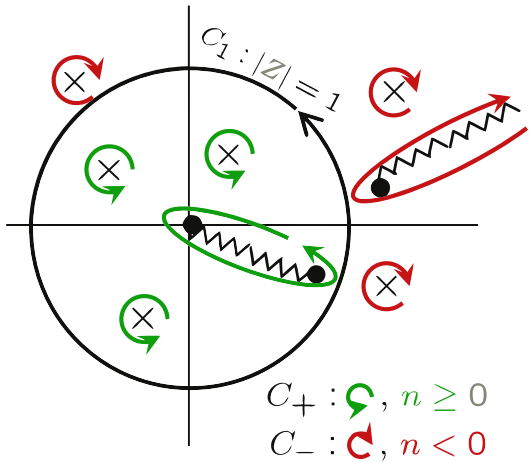}
  \caption{The integration contours for the IZT. Poles are marked by $\times$, branch points by $\bullet$, and branch cuts by wiggly lines. Singularities inside (outside) the unit circle contribute to $n\ge 0$ ($n<0$).}\label{fig1App_ZT}
\end{figure}

The double-sided $Z$ Transform (ZT) of a \emph{bounded} vector or matrix series, say $\bb{q}_n$, is obtained by applying the conventional (scalar-series) $Z$ transform to each of the entries.
Hence
\begin{equation}
\bb{\hat{q}}(Z)=\sum_n^\infty \bb{q}_n\, Z^{-n}\label{eqApp_ZT1}
\end{equation}
and the transform of a matrix $\underline{\bf D}_{\, n}$ is obtained similarly.
The series region of convergence (ROC) is  a ring that contains the unit circle $C_1:\,\, \abs{Z}=1$.
The Inverse ZT (IZT) is given by
\begin{equation}
\bb{q}_n=\frac{1}{2\pi i}\oint_{C_\pm} \bb{\hat{q}}(Z)\, Z^{n-1} dZ. \label{eqApp_ZT2}
\end{equation}
The original integration contour should reside within the ROC, and encircle the origin in a counter clockwise direction; $C_1$ is an appropriate path, as shown in Fig.~\ref{fig1App_ZT}. For detailed mathematical discussion the reader is referred e.g.~to \cite{ZTBook1} where it is termed as the two-sided transform.

To enhance physical insight, however, we shall replace the original contour with integrations around singularities (poles and branch cuts) of the inverse transform kernel.
Thus, $C_1$ is replaced with $C_{\pm}$.
For observation points located at $n\ge 0$ the integration contour $C_\pm=C_+$ encircles all the singularities \emph{within} $C_1$ in the complex $Z$ plane in a counter clockwise direction. The contour $C_\pm=C_-$ used for $n<0$ encircles all the singularities \emph{external} to $C_1$ in the complex $Z$ plane in a clock-wise direction. The contours are shown in Fig.~\ref{fig1App_ZT}.
The contributions of the different singularities  may readily be used to discern between various wave species.
Poles located on (off) the unit circle corresponds to propagating (radiation) modes  whereas branch-points and cuts correspond to continuous spectrum (CS) waves. A detailed discussion of the application of the ZT to propagation in chains of dielectric particles and the aforementioned association of physical phenomena (chain waves) with mathematical singularities can be found in \cite{HadadSteinbergGreen}.

\section{Reciprocity}\label{App_Recip}
To discuss reciprocity, it is best if we start from the general statement of the reciprocity theorem \cite{RumseyPaper}
\begin{multline}
\int_V\left(\bb{E}_1\cdot\bb{J}_2-\bb{H}_1\cdot\bb{J}_{m2}\right)=\\
\int_V\left(\bb{E}_2\cdot\bb{J}_1-\bb{H}_2\cdot\bb{J}_{m1}\right)
\label{eq25}
\end{multline}
where $\bb{E}_{1,2}$ and $\bb{H}_{1,2}$ are the fields generated by electric and magnetic sources $\bb{J}_{1,2}$ and $\bb{J}_{m1,m2}$ and all vectors are regarded as column vectors. We use the pre-defined unified notations and obtain
\begin{multline}
\int_V\left[\bb{J}^T_2,-\bb{J}_{m2}^T/\eta\right]\bb{F}_1=\\
\int_V\left[\bb{J}^T_1,-\bb{J}_{m1}^T/\eta\right]\bb{F}_2
\label{eq26}
\end{multline}
We would like to examine this relation for the point sources
\begin{equation}
\bb{J}=-i\omega\bb{p}\delta(\bb{r}-\bb{r}')\;,\;\bb{J}_m=-i\omega\bb{m}\delta(\bb{r}-\bb{r}')
\label{eq27}
\end{equation}
which yields the equation
\begin{multline}
\left[\bb{p}_2^T,-\bb{m}_2^T\right]{\bf G}(\bb{r}-\bb{r}')
\left[
\begin{matrix}
\bb{p}_1 \\
\bb{m}_1
\end{matrix}\right]=\\
\left[\bb{p}_1^T,-\bb{m}_1^T\right]{\bf G}(\bb{r}'-\bb{r})
\left[
\begin{matrix}
\bb{p}_2 \\
\bb{m}_2
\end{matrix}\right]
\end{multline}
Other then stating the condition for reciprocity (which can be translated to properties for $\bf G$, for example \cite{BIbook}, Eqs. 2.143-2.145) this states another important fact - when switching the role of a general Huygens point dipole from source to observer, one must \emph{flip} the direction of the magnetic dipole. This is because the electric current is a vector, whereas the magnetic current is a pseudovector \cite{JacksonBook}.

In the context of our problem, the switching $\bb{J}_m\mapsto -\bb{J}_m$ accounts precisely for the passage from the chains eigenstate $q=1$ to $q=2$ (see table \ref{Table1}), hence reciprocity is satisfied.


\begin{thebibliography}{1}

\bibitem{Markel1}
V.~A.~Markel,
``Coupled-dipole Approach to Scattering of Light from a One-dimensional Periodic Dipole Structure,''
\emph{J.~Mod.~Opt.}, {\bf 40}(11) 2281-2291, DOI:
10.1080/09500349314552291 (1993).

\bibitem{Viitanen}
S.~A.~Tretyakov and A.~J.~Viitanen,
``Line of periodically arranged passive dipole scatterers,''
\emph{Electr. Eng. Rev.}, {\bf 82} 353 (2000)

\bibitem{TretyakovChain}
C.~R.~Simovski, A.~J.~Viitanen, and S.~A.~Tretyakov, ``Resonator modes in chains of silver spheres and its possible application,'' \emph{Phys.~Rev.~E} {\bf 72}, 066606 (2005).

\bibitem{EnghetaChain}
A.~Alu and N.~Engheta, ``Theory of linear chains of metamaterial/plasmonic particles as subdiffraction optical nanotransmission
lines,'' \emph{Phys.~Rev. B} {\bf 74}, 205436 (2006).

\bibitem{HadadSteinbergGreen}
Y.~Hadad, Ben~Z.~Steinberg ``Green's function theory for infinite and semi-infinite particle chains,'' \emph{Phys.~Rev.~B} {\bf 84}, 125402 (2011).

\bibitem{CapolinoChain}
S.~Campione, S.~Steshenko, and F.~Capolino,
``Complex bound and leaky modes in chains of plasmonic nanospheres,"
\emph{Optics Express}, {\bf 19}(19), 18345-18363 (2011).

\bibitem{Vitaliy1}
D.~V.~Orden, Y.~Fainman, and V.~Lomakin,
``Optical waves on nanoparticle chains coupled with surfaces,''
\emph{Opt. Lett.}, {\bf 34}(4) 422 (2009)


\bibitem{HadadSteinbergPRL}
Y.~Hadad and Ben Z.~Steinberg,
``Magnetized spiral chains of plasmonic ellipsoids for one-way optical waveguides,''
\emph{Phys.~Rev.~Lett.}, {\bf 105} 233904 (2010)

\bibitem{MazorSteinberg_LongChir}
Y.~Mazor and Ben Z.~Steinberg,
``Longitudinal chirality, enhanced nonreciprocity, and nanoscale planar one-way plasmonic guiding,''
\emph{Phys.~Rev.~B}, {\bf 86} 045120 (2012)

\bibitem{Vitaliy2}
D.V. Orden, Y. Fainman, and V. Lomakin,
``Electromagnetic waves on twisted linear arrays,''
\emph{Opt. Lett.}, {\bf 35} 2579 (2010)

\bibitem{HadadSteinbergOpEx}
Y.~Hadad and Ben Z.~Steinberg,
``One way optical waveguides for matched non-reciprocal nanoantennas with dynamic beam scanning functionality,''
\emph{Optics Express}, {\bf 21}(S1) A77 (2013)


\bibitem{Tretyakov1}  
Y.~Ra'di, V.~S.~Asadchy, and S.~Tretyakov,
``Tailoring Reflections from Thin Composite Metamirrors,''
\emph{IEEE Trans.~Ant.~Propag.}, {\bf 62}(7), 3749-3760 (2014).

\bibitem{Tretyakov2} 
M.~Yazdi, M.~Albooyeh, R.~Alaee, V.~Asadchy, N.~Komjani, C.Rockstuhl, C.~R.~Simovski, and S.~Tretyakov,
``A Bianisotropic Metasurface with Resonant Asymmetric Absorption,''
\emph{IEEE Trans.~Ant.~Propag.}, {\bf 63}(7), 3004-3015 (2015).

\bibitem{TretyakovReview} 
Y.~Ra'di, C.~R.~Simovski, and S.~A.~Tretyakov,
``Thin Perfect Absorbers for Electromagnetic Waves: Theory, Design, and Realizations,''
\emph{Phys.~Rev.~Applied}, {\bf 3}, 037001 (2015)

\bibitem{TretyakovPRX2015} 
V.~S.~Asadchy, I.~A.~Faniayeu, Y.~Ra'di, S.~A.~Khakhomov, I.~V.~Semchenko, and S.~A.~Tretyakov,
``Broadband Reflectionless Metasheets: Frequency-Selective Transmission and Perfect Absorption,''
\emph{Phys.~Rev.~X}, {\bf 5}, 031005 (2015)

\bibitem{TretyakovOneWaySheets} 
Y.~Ra'di, V.~S.~Asadchy, and S.~A.~Tretyakov,
``One-way transparent sheets,''
\emph{Phys.~Rev.~B} {\bf 89}, 075109 (2014)

\bibitem{Capolino_MagnetoElectric3D}
S.~Campione and F.~Capolino,
``Electromagnetic coupling and array packing induce exchange of dominance on complex modes in 3D periodic arrays of spheres with large permittivity,''
\emph{J.~Opt.~Soc.~Am.~B}, {\bf 33}(2) 261-270 (2016)


\bibitem{AluHomog1}
Xing-Xiang Liu and A.~Alu,
``Homogenization of quasi-isotropic metamaterials composed by
dense arrays of magnetodielectric spheres,''
\emph{Metamaterials}, {\bf 5} 56-63 (2011)

\bibitem{AluHomog2}
Xing-Xiang Liu, J.~W.~Massey, Ming-Feng Wu, K.~T.~Kim, R.~A.~Shore, A.~E.~Yilmaz, and A.~Alu,
``Homogenization of three-dimensional metamaterial objects and validation by a fast surface-integral equation solver,''
\emph{Optics Express}, {\bf 21}(18) 21714-21727 (2013)

\bibitem{Mag1}
A.~Franchini, V.~Bortolani, and R.~F.~Wallis, 
``Interaction of an external impurity with the surface intrinsic mode in a Heisenberg chain,''
\emph{Phys.~Rev.~B}, {\bf 73}, 054412 (2006).

\bibitem{Mag2}
R.~Zivieri,
``Metamaterial Properties of One-Dimensional and Two-Dimensional Magnonic Crystals,''
\emph{Solid State Physics}, {\bf 63} 151-215 (2012).

\bibitem{FemiusUnits}
I.~Sersic, C.~Tuambilangana, T.~Kampfrath, and A.~F.~Koenderink,
``Magnetoelectric point scattering theory for metamaterial scatterers,''
\emph{Phys.~Rev.~B} {\bf 83}, 245102 (2011).

\bibitem{SihvolaSphere}
D.~C.~Tzarouchis, P.~Yla-Oijala, and A.~Sihvola,
``Unveiling the scattering behavior of small spheres,''
\emph{Phys.~Rev.~B} {\bf 94}, 140301 (2016).

\bibitem{Campione_Balanced}
S.~Campione, L.~I.~Basilio, L.~K.~Warne, and M.~B.~Sinclair,
``Tailoring dielectric resonator geometries for directional scattering and Huygens' metasurfaces,''
\emph{Optics Express}, {\bf 23}(3) 2293-2307 DOI:10.1364/OE.23.002293 (2015)



\bibitem{ZTBook1}
Eliahu I. Jury, \emph{Theory and Applications of the Z-Transform Method} (Wiley, New York, 1964).

\bibitem{ZTBook2}
Saber N. Elaydi, \emph{An Introduction to Difference Equations} 3rd ed.
(Springer, New York, 2005).

\bibitem{ZT_WasylWasyl}
W.~Wasylkiwskyj,
``Mutual Coupling Effects in Sem-Infinite Arrays,''
\emph{IEEE Trans.~Ant.~Propag.}, {\bf 21}(3), 277-285 (1973).

\bibitem{ZT_SIAM}
C.~M.~Linton and P.A.~Martin,
``Semi-Infinite Arrays of Isotropic Point Scatteres - A Unified Approach,''
\emph{SIAM J.~Appl. Math.}, {bf 64}(3), 1035-1056 (2004).

\bibitem{CapolinoZTWH}
F.~Capolino and M.~Albani,
``Truncation effects in semi-infinite periodic array of thin strips: A discrete Wiener-Hopf formulation,''
\emph{Radio Science} {\bf 44}, RS2S91, doi:10.1029/2007RS003821 (2009).


\bibitem{PolylogBook}
L.~Lewin, \emph{Polylogarithms and Associated Functions}, Elsevier, New York, 1981.

\bibitem{RumseyPaper}
V.H.~Rumsey, ``Reaction Concept in Electromagnetic Theory,'' \emph{Phys.~Rev.} {\bf 94}, 6, 1483-1491, (1954)

\bibitem{BIbook}
I.V.~Lindell, A.H.~Sihvola, S.A.~Tretyakov, A.J.~Viitanen, \emph{Electrimagnetic waves in Chiral and Bi-Isotropic media}, Artech House, 1994.

\bibitem{JacksonBook}
J.D.~Jackson, \emph{Classical Electrodynamics, 3rd Edition}, Wiley, New York, (1999).

\end{thebibliography}
\end{document}